\documentclass[11pt,a4paper]{article}
\usepackage{graphicx}
\usepackage{amsmath}
\usepackage{caption}
\usepackage{authblk}
\usepackage{geometry}
\usepackage[superscript,biblabel]{cite}
\title{Experimental demonstration of single-shot quantum and classical signal transmission on single wavelength optical pulse}

\author[1,2,*]{Rupesh Kumar}
\author[2]{Adrian Wonfor}
\author[2]{Richard Penty}
\author[1]{Tim Spiller}
\author[2]{Ian White}
\affil[1]{Quantum Communications Hub, Department of Physics, University of York}
\affil[2]{Centre for Photonic Systems, Department of Engineering, University of Cambridge}
\affil[*]{rupesh.kumar@york.ac.uk}

\date{}

\begin{document}

\maketitle

\textbf{Advances in highly sensitive detection techniques for classical coherent communication systems have reduced the received signal power requirements to a few photons per bit. At this level one can take advantage of the quantum noise to create secure communication, using continuous variable quantum key distribution (CV-QKD). In this work therefore we embed CV-QKD signals within classical signals and transmit classical data and secure keys simultaneously over 25km of optical fibre.  This is achieved by using a novel coherent displacement state generator, which has the potential for being used in a wide range of quantum optical experiments.  This approach removes the need for separate channels for quantum communication systems and  allows reduced system bandwidth for a given communications specification.  This demonstration therefore demonstrates a way of implementing direct quantum physical layer security within a conventional classical communications system, offering a major advance in term of practical and low cost implementation.}

The increasing demand for data communication bandwidth is increasingly being met by  classical coherent communications using complex modulation formats. Here data transmission capacity is increased substantially over on-off keying (OOK) by using  muti-level modulation of the amplitude and phase- jointly referred as quadratures, of the light{\cite{Kikuchi2016}}. 
At the lowest signal levels, transmission of messages are  disrupted  by the  quantum uncertainty of the signal properties and their measurements.  However, quantum uncertainty can also be used  for secure cryptographic key distribution{\cite{Bennett1984}}. At the quantum level, the on-off single photon detection technique enables extraction of secure key from single photon pulses or their superpositions. Similarly, vacuum  noise sensitive coherent  detection  can extract keys from the quadratures of low light signals. The former is referred to as discrete variable (DV) QKD and the latter is continuous variable (CV) QKD{\cite{Jouguet2013a}}. In both systems, it is necessary to use auxiliary intense signals primarily  to assist the  signal measurement and subsequent secure key generation{\cite{Jouguet2013a,Stucki2011,Choi2011,Roberts2017,Ikuta2018}}. 

Simultaneous transmission and detection of classical and quantum signals  in a single signal is not feasible in DV-QKD systems as it is difficult to discriminate the quantum  and classical data from the  single photon detector outcomes.  On the other hand, CV-QKD systems\cite{Jouguet2012a} can be modified to incorporate simultaneous encoding, transmission,  detection and decoding   of quantum and classical signals in the same mode. At intermediate signal strengths, where the classical coherent communication signal exhibits  quantum uncertainty,   we can simultaneously transmit    a quantum key as well as  classical message in a single mode{\cite{Bing2016,Kumar2018}}. As the intensity of the signal is comparatively higher, there is a reduced need for intense auxiliary signals.  This is important  in wavelength division multiplexing (WDM)   or time multiplexing QKD networks\cite{Frohlich2013,Townsend1997,Choi2011,Kumar2015} where otherwise  multiple channels need to be allocated for a single QKD link. This is particularly difficult  in networks with dense conventional data traffic{\cite{Bahrani2018}}. 

 CV-QKD uses quadrature modulated coherent states with very low numbers of photon per signal, e.g. based on modulation variance, 0.1 - 100 photons on average at the transmitter{\cite{Jouguet2011}}, to establish secure keys between two users- namely Alice  and Bob.  Being extremely low in intensity,  coherent signals exhibit shot-noise (vacuum noise)   uncertainty- which imparts  security {\cite{Grosshans2002}}, in their quadrature values and thus requires shot-noise limited coherent detectors. This stringent requirement  is absent in classical coherent detection which uses comparatively intense signals{\cite{Kikuchi2016}}. 
 
In order to transmit both CV-QKD signal and classical data signal on a single wavelength pulse, we require a signal that shows quantum uncertainty- for secrecy, as well as elevated intensity- for deterministic discrimination of classical data. Displacing weak CV-QKD signals to comparatively bright coherent ones meets this requirement{\cite{Bing2016}}.  Here we first describe the  technique used for quantum state displacement at Alice,   experimental setup for generating CV-QKD signals with displaced signals  and  finally state measurement at Bob for secure key distribution. 

\section*{Result and Discussion}

\begin{figure}[htb!]
 \includegraphics[width=0.99\textwidth]{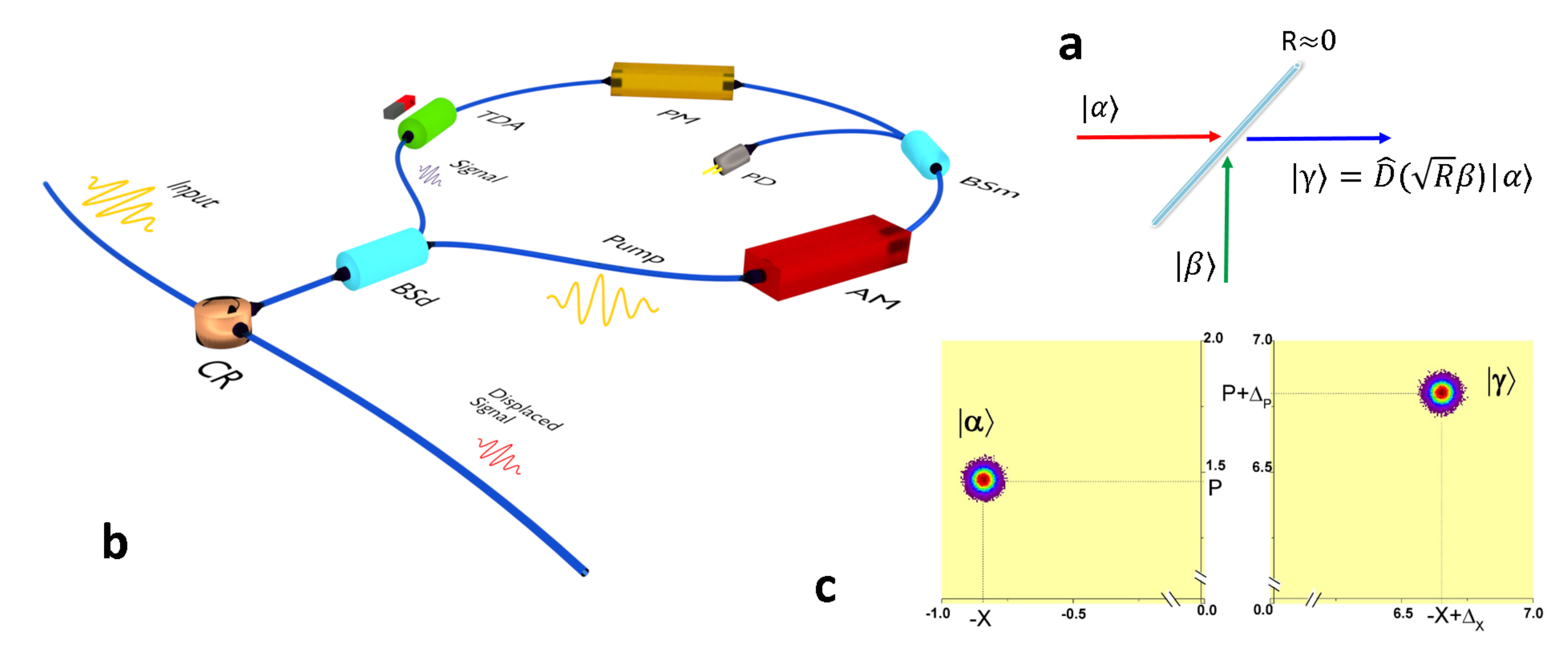}
\caption{\textbf{Displaced signal generation- concept and realization. a,} Displacement operation using a  highly asymmetric beam splitter of transmissivity  $T\rightarrow 1 ( R = 1-T \approx 0)$.  Mixing a strong pump $|\beta\rangle$ with the coherent state $|\alpha\rangle$ at the beam splitter displaces  the signal by $\sqrt{R}\beta$. \textbf{b,} Displaced signal generator (DSG) setup. Polarization maintaining fibres  and components are used to  lock the polarization of the signal and  the pump. \textbf{c,} Measured Signal before and after displacement in phase space with respect to the local oscillator's reference frame. The quadrature components $\{-X, P\}$ of the initial state $|\alpha\rangle$ is shifted to $\{-X+\Delta_X, P+\Delta_P\}$ of the final state$|\gamma\rangle =$  $j|\alpha+\sqrt{R}\beta\rangle$  where, $j=\exp(\alpha \sqrt{R}\beta^* - \alpha^* \sqrt{R}\beta)/2$ is the global phase that sets the direction of displacement with respect to the local oscillator.  Here, $\Delta_X$ and $\Delta_P$ are the respective quadrature components of the displacement $\Delta=\sqrt{R}\beta$. Here $R=0.01$ and intensity of the pump $|\beta|^2\approx$  6100 photons per pulse. The colour map shows the measured  signal and displaced signal  with shot-noise uncertainty. Quadrature values are expressed in $\sqrt{N_0}$, where $N_0$ is the shot noise variance. Please refer the text for the expansion of abbreviations  used in  the setup. }
\label{DSG}     
\end{figure}

The displacement operator $\hat{D}(\Delta)$ is an important tool in quantum optics that displaces any input state  $|\alpha\rangle$ to any other state  $|\beta\rangle$ in the phase space{\cite{Cook2007,Ferdinand2017,Izumi2018}}.
As a good practical approximation, mixing a coherent state with a strong pump at a  highly transmitive  beam splitter can generate a displaced coherent state{\cite{Paris1996}}. For proper displacement operation, it is necessary to lock the phase of the pump  to the signal. We  have devised a novel way of achieving this based on a  Sagnac interferometer which is inherently stable.   As shown in Fig.\ref{DSG}, the Sagnac loop incorporates a highly asymmetric beam splitter(BS$_d$) with a 99/1 split ratio. The  input pulse to  BS$_d$ is  split into two- a strong pump  propagating in the clockwise  direction and weak signal  in the anti-clockwise direction. Amplitude (AM) and phase (PM) modulators are used to prepare the initial coherent state $|\alpha\rangle$ and  the pump. 

After propagating through the Sagnac loop, both the pump and signal meet at the BS$_d$, which strongly reflects ($99\%$) the pump and  at the same time largely   transmits ($99\%$)  the signal. The pump  displaces the signal  in phase space according to the relative phase between the pump and signal and the splitting ratio of BS$_d${\cite{Paris1996}}. The input port of  BS$_d$ now acts as the output port for the displaced state $|\gamma\rangle$. The signal amplitude is monitored using a 99/1 beam splitter (BS$_m$) and photodiode (PD). The tunable directional attenuator (TDA) which is an optical Faraday isolator with tunable external magnetic field, that provides -3dB to -30dB of attenuation in the signal.   The circulator (CR) serves as the access port of the Sagnac loop. We define this setup as the displaced signal generator (DSG). It should be noted that  in the DSG setup the pump and  the signal pulses can independently  be modulated by all available amplitude and phase  levels allowed by the digital to analog converters. As a result, the DSG has greater freedom to generate different quantum and classical modulation formats.
 

\begin{figure}[htb!]
 \includegraphics[width=0.99\textwidth]{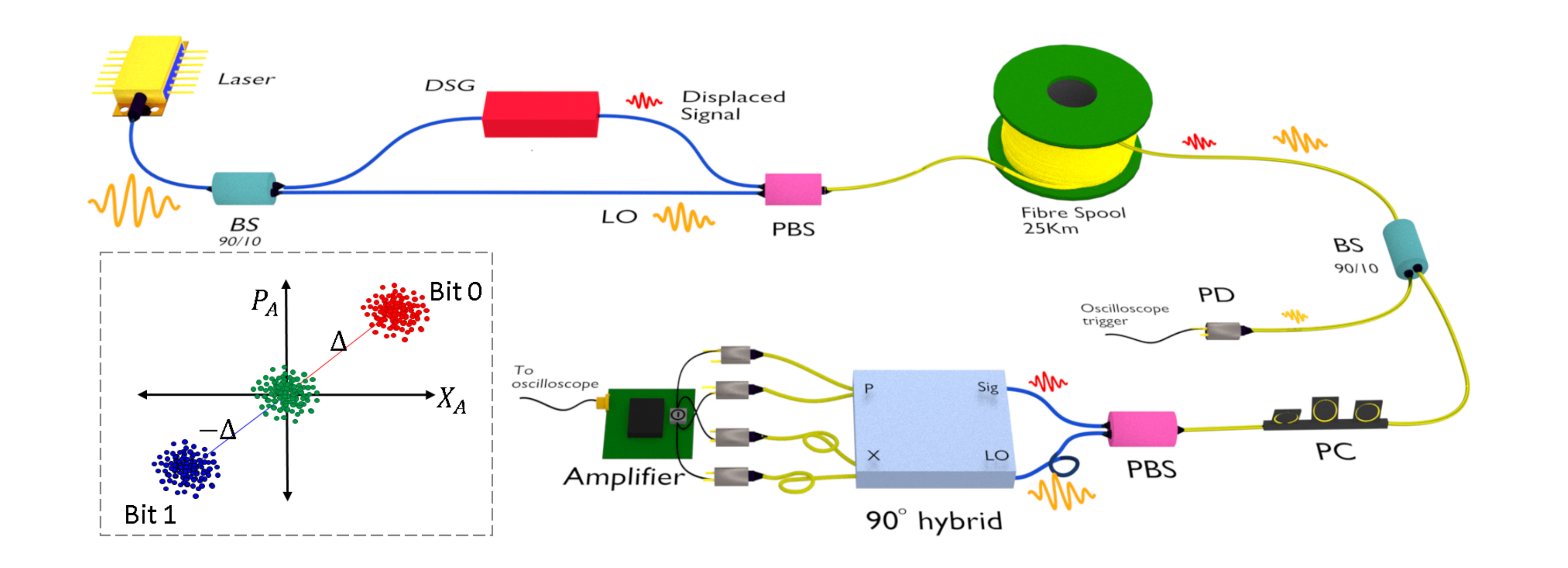}
\caption{\textbf{Experimental setup}. The displaced coherent state generator is integrated into a conventional CV-QKD system with hetrodyne detection. The LO and displaced signals are time (75ns) and polarization  multiplexed at Alice and demultiplexed at Bob.  The blue fibre is polarization maintaining and the yellow is SMF. A single low noise amplifier is used for X and P quadrature measurements. The scheme for Gaussian modulated displaced signals (red and blue) with BPSK modulation, at Alice is shown in inset. The green dots represents conventional Gaussian modulated signals that are shown for comparison.  More details of the experimental setup is given in method. }
\label{EXPsetup}     
\end{figure}

To achieve single-shot quantum and classical  communication, a fibre based CV-QKD system  has been constructed, as shown in Fig.\ref{EXPsetup}, with a DSG at Alice and  a shot noise limited coherent receiver to perform heterodyne detection at Bob. A 25ns pulse at 1556nm is fed into the DSG,  where  the signal $|\alpha\rangle_A$  is  amplitude and phase modulated in such a way that the quadrature, $\{X_A, P_A\}$, of the signal before the displacement follows a Gaussian distribution $ \mathcal{N}\{0, V_A\}$ with  zero mean and variance $V_A$\cite{Grosshans2002}, as in the case of conventional CV-QKD systems. We set the variance to $5N_0$ which is optimal for 25km transmission distance.  Here, $N_0$ is the shot noise variance.   The classical data is encoded on the phase of the pump such that the direction of displacement follows different classical modulation format. 
The displaced signal with quadrature $\{X_A+\Delta_{x}, P_A+\Delta_{p}\}$, is then sent to Bob. The distribution of the quadratures is now modified to $ \mathcal{N}\{\Delta_{x(p)}, V_A\}$, i e. the mean of the Gaussian distribution is shifted from zero to respective displaced levels, and the magnitude of $\Delta$ is set to $13\sqrt{N_0}$ at Bob. 

At the receiver, Bob measures the quadrature, $\{X_{B\Delta_{x}},P_{B\Delta_{p}}\}$, of the displaced signals with a shot noise limited heterodyne detector. From the measurement outcomes, the direction of displacement has been estimated from which classical data bits are inferred.  By offsetting the mean to zero, the quadrature of the Gaussian modulated signals can be retrieved.  This scheme has been to be proven secure as it implements the conventional Gaussian modulated protocol\cite{Bing2016}. The displacement does not affect, in principle, the signal variance, $V_A$ and channel parameter estimation-transmittance $T$ and excess noise $\xi$, as it only affects the mean of  measurement outcomes{\cite{Bing2016}}. However, estimating the magnitude of the displacement from each of Bob$\textquoteright$s measurement outcomes and deducing respective quadrature values  increases the excess noise{\cite{Adrien2017}}.  Here, the displacement is estimated  independently, from the pilot signals that are used for phase drift compensation.  The phase reference- local oscillator (LO), for the quadrature measurement at the Bob, is sent from Alice along with each displaced signals. \\

\begin{figure}[htb!]
 \includegraphics[width=1\textwidth]{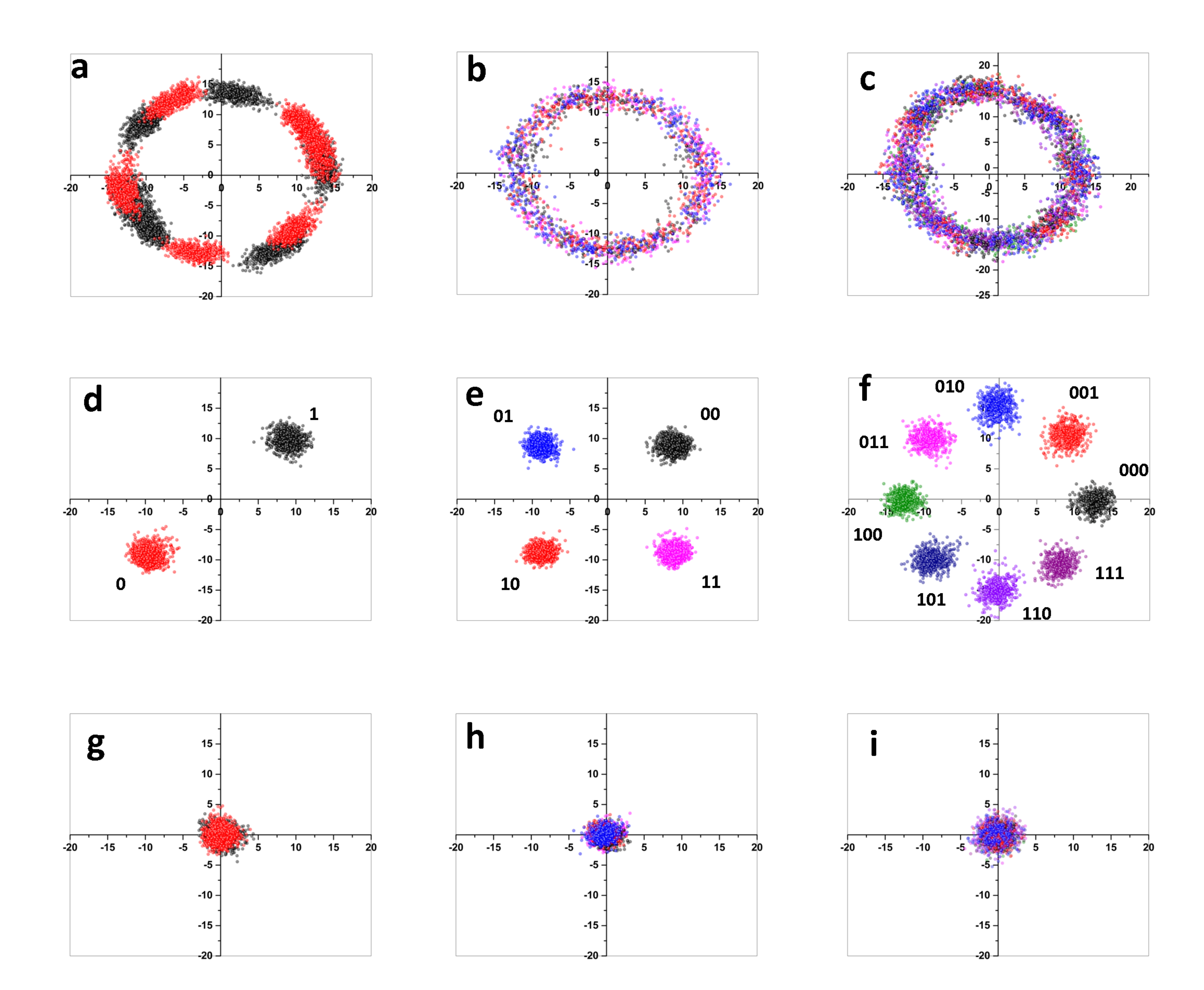}
\caption{\textbf{Experimental result}. Results are plotted with $X$ vs $P$ quadrature measured at Bob. \textbf{a,b,c}- respectively BPSK, QPSK and 8PSK, show the received signal quadratures which are randomly phase rotated due to relative phase drift between displaced signals and LO. \textbf{d,e,f} are the phase corrected displaced signals  which clearly show the transmitted classical modulation format. The bit values associated with each group of signals are publicly agreed by Alice and Bob  prior to the transmission of classical data. \textbf{g,h,i} are the deduced Gaussian modulated continuous variable data from the displaced signals.}
\label{Result}     
\end{figure}
Fig.\ref{Result} shows the experimental results for different classical modulation patterns.  An arbitrary relative phase drift between the displaced signals and LO results in the rotation of the quadrature values in phase space. The phase drift is estimated from pilot signals and applied on the received signals. The classical bit values are inferred from the phase drift corrected quadrature values.  In the BPSK modulation scheme, Fig.\ref{Result}d, 1.2Mb/s data can be transmitted at a 2MHz clock  rate. The effective rate is reduced to 1.2MHz due to a $9\%$ pilot signal allocation and $30\%$ to shot noise measurement. The average excess noise estimated from quantum signal quadratures is $0.055N_0$ with a transmittance T = 0.28. The estimated secure key rate is 30kb/s. 
The estimated Q factor of the BPSK signal is 9.6 and leads to an expected error free operation with a bit error rate (BER) of $10^{-20}$. In the QPSK modulation scheme, Fig.\ref{Result}e, the overall classical data rate is 2.9Mb/s with a 25kb/s  estimated secure key rate. The excess noise is found to be $0.047N_0$ with T = 0.23. The Q factor of the classical communication system is 8.8 indicating a BER of $10^{-17}$. Using the 8PSK modulation, Fig.\ref{Result}f, a classical data rate of 4.9Mb/s is achieved with a 25kb/s estimated secure key rate at $0.053N_0$ excess noise and T = 0.24. The Q factor is reduced to 4.6 and the BER is $\approx 10^{-6}$, which can be made to operate error free either by using large displacement values  or  by using forward error correction  methods.  Future work should consider   more complex classical modulation formats, such as 16QAM and 32QAM.
 
To conclude,  this paper reports the demonstration  of simultaneous  quantum and  classical data transfer on  a single signal. This technique enables direct monitoring of eavesdropping on classical communications where classical signals are combined  with quantum signals using DSG{\cite{Yupeng2018}} which may have a strong commercial interest as it provides physical layer security using standard telecommunication equipment. Especially using 10G/100G classical coherent communication systems where one can fairly estimate channel parameters of  CV-QKD but it is hard to generate secure key due to poor performance of error correction and privacy amplification at high data rate{\cite{Jouguet2011}}. Aside from these promising applications, the DSG may find applications in other quantum optics experiments that require stable realizations of a displacement operator{\cite{Cook2007,Ferdinand2017,Izumi2018}}.  Finally, the displaced signals can be detected with Local Local Oscillator based coherent detection, however the excess noise  due to phase estimation and phase drift may significantly affect the transmission distance, key rate and BER values. However, we anticipate this  can be demonstrated in our future work.

\section*{Methods}
\textbf{Experimental details}\\
In our experimental setup, at Alice, a 25ns pulse at a 2MHz repetition rate and wavelength of 1556nm is split by an asymmetric beam splitter.  The stronger pulse serves as the LO for Bob$\textquoteright$s measurements. The weaker pulse is sent to  the DSG. The displaced signal and LO are polarization multiplexed by a polarization beam splitter (PBS). They are time delayed by 75ns owing to the difference in path length inside Alice. The displaced signal and LO pulse are sent to Bob through a 25km SMF fibre channel. The fibre channel scrambles the polarization of the pulses during the transit. 
At Bob, a polarization controller (PC) corrects the polarization offset of the displaced signal and LO pulses. Another PBS demultiplexes the incoming pulses into a short path for the signal and a long path for the LO. A fibre delay of 75ns in the LO path at Bob compensates for the 75ns pulse time delay at the transmitter. The signal pulse and LO pulse are then fed into a $90^o$ hybrid optical device (Kylia COH24). The P quadrature output of the hybrid is connected to a pair of PIN photodiodes. The subtracted current is amplified by a 10MHz bandwidth low-noise amplifier based on an Amptek A250 charge amplifier. The same amplifier circuit is used for the X quadrature measurement where the output of the $90^o$ hybrid is optically delayed by 125ns and connected to another pair of PIN diodes. This increases the overall detection bandwidth to 4MHz. Each quadrature measurement is independently balanced to set the mean of the amplifier output to zero volts. The data is acquired by a 4GHz bandwidth, 10bit, real time oscilloscope and post processed in a computer in real time. The oscilloscope is synchronized with the LO signal using a $10\%$ fibre tap after the 25km channel. The quadrature values are measured in the unit of $mV$ and normalized to $\sqrt{N_0}$. 



\noindent \textbf{Phase correction and displacement estimation}\\
The phase drift  relative to displaced signal and LO is estimated using pilot signals. The transmission is sliced into  packets  of 400 signals that contain pilot signals with multiple phase pattern. The amplitude of the pilot signals are set to equal to the magnitude of the displacement. Each phase pattern set comprises of three signals  with $2\pi/3$ mutually relative phase difference. The magnitude of the displacement is estimated as $\Delta = \frac{1}{k} \sum_{j=1}^{k}\sqrt{\frac{2}{3}\sum_{i=1}^{3}(u_{ij} - m_j)^2}$, where $u_{ij}$ is the measured quadrature of the pilot signals in pattern set $j$ and  $m_j = \frac{1}{3}\sum_{i}^{3}u_{ij}$  is the mean. More than $k=10^7$ pattern sets are used to estimate the displacement. The phase drift in each packet is  estimated as  $\theta_{d} = \sum_{j=1}^{p}  \tan^{-1}[ (u_{2j} -u_{3j})/(\sqrt{3}(2u_{1j}-u_{2j} -u_{3j}))]+\theta_{j}^{offset}$, with $p =24$ is the total number of pattern sets  measured in each packet, in both quadratures and $\theta_{j}^{offset}$ is the offset from $0^o$. We use exponential averaging over  successive $\theta_{d}$ to increase the precision. 

\noindent\textbf{CV-QKD parameters and key estimation }\\
The channel parameters, transmittance  $T$ and excess noise  $\xi$, are estimated  by making use of following equations: $\langle X_A X_B\rangle= \sqrt{\gamma\eta T}V_A$ and $V_B=\gamma\eta T V_A+ N_0+\gamma\eta T \xi +V_{ele}$. These equations are true for parameter estimation from P quadrature  measurements. Here $\eta = 0.62$  is the efficiency of Bob and $V_{ele}=0.01N_0$ is the variance of the electronic noise variance, all are independently estimated. $\gamma=1$ for homodyne detection and $\gamma=1/2$  for heterodyne detection (in our case). $X_A$ and $X_B$ are the quadrature  and $V_A$ and $V_B$ are the respective variance,  without displacement at Alice and Bob, respectively. The parameters are estimated from $10^8$ samples of both X and P quadratures.  

The secret key rate, under collecive attack, in reverse reconciliation is  \cite{Lodewyck2007}  $K= R(\beta I_{AB}-\chi_{BE})$ where $R$ is the effective signal rate, $\beta$ is the reconciliation efficiency,  $I_{AB}=  \log_2\left( \frac{V+\chi_{tot}}{1+\chi_{tot}}\right)$ is the mutual information between Alice and Bob in heterodyne detection scheme. Here $V= V_A +1$ is the signal variance that includes Gaussian modulation by Alice,$V_A$ and shot noise variance normalized to one. $\chi_{BE} = \sum_{i=1}^2 G\left(\frac{\lambda_i -1}{2} \right)- \sum_{i=3}^5 G\left(\frac{\lambda_i -1}{2} \right)$ is the  maximum information gain by Eve under collective attack. Where  $G(x) = (x+1)\log_2(x+1)-x\log_2 x$ and $\lambda_{1,2}^2 = \frac{1}{2}\left(A\pm\sqrt{A^2-4B}\right)$ and  $\lambda_{3,4}^2 = \frac{1}{2}\left(C\pm\sqrt{C^2-4D}\right)$ and $\lambda_5 =1$.  The term $\chi_{tot}$ is the total noise at the output of Alice and expressed as $\chi_{tot}=\chi_{line}+ \frac{\chi_{het}}{T}$.   The channel noise $\chi_{line}= \frac{1}{T} -1 +\xi$  and the detection noise $\chi_{het} =  (2 + 2v_{ele} -\eta)/\eta$.    Here, $ A = V^2(1-2T)+2T+T^2(V+\chi_{line})^2$, $
 B= T^2(V\chi_{line}+1)^2 $, $
C = \frac{1}{(T(V+\chi_{tot}))^2}[ A\chi_{het}^2 +B+1+ 
   2\chi_{het}(V\sqrt{B}+T(V+\chi_{line}))+2T(V^2-1)]$,  and $
D =  (\frac{V+\sqrt{B}\chi_{het}}{T(V+\chi_{tot}})^2$.

\noindent\textbf{Bit error rate in classical coherent communication}\\
The Bit Error Rate (BER) of the detected signals is estimated using BER= 0.5 erfc$(Q/\sqrt{2})$, where erfc(*) denotes the complementary error function. The Q factor defines as $|d_0 -d_1|/(\sigma_0+\sigma_1)$ , where $d_i$ is the mean of the respective signal levels and $\sigma_i$ is the standard deviation- mainly due to Gaussian signal modulation.

\bibliography{cvref}

\section*{Acknowledgement}
We wish to acknowledge the funding from the UK Engineering and Physical Sciences Research Council (EPSRC) through UK Quantum Technology Hub for Quantum Communications Technologies, Grant no. EP/M013472/1.
\section*{Authors contribution}
R.K performed the experiment and analysed the data.  A.W, R.P and I.W provide scientific expertise in classical communication. T.S and I.W supervised the project. R.K wrote the manuscript with contributions from all the authors.
\section*{Additional information}
Correspondence and
requests for materials should be addressed to R.K
\section*{Competing  interests}
The authors declare no competing interests.

\end{document}